\documentclass[twocolumn,aps,showpacs,prl,epsf,graphics,psfig]{revtex4}
\usepackage{graphicx}

\begin{document}
\title{Tailoring Magnetism in Quantum Dots}

\author{Ramin M. Abolfath$^{1,2}$}
\author{Pawel Hawrylak$^2$}
\author{Igor \v{Z}uti\'c$^1$}

\affiliation{
$^1$Department of Physics, State University of New York at Buffalo,
Buffalo, New York 14260, USA \\
$^2$Institute for Microstructural Sciences,
National Research Council of Canada, Ottawa, K1A 0R6, Canada 
}

\date{\today}

\begin{abstract}
We study magnetism in magnetically doped quantum dots as a function of 
confining potential, particle numbers, temperature, and strength of 
Coulomb interactions. 
We explore possibility of tailoring magnetism 
by controlling the electron-electron Coulomb interaction, without changing 
the number of particles. The interplay of strong Coulomb interactions and 
quantum confinement leads to enhanced inhomogeneous magnetization which 
persist at higher temperatures than in the non-interacting case.
The temperature of the onset of magnetization can be controlled by
changing the number of particles as well as by modifying the quantum 
confinement and the strength of Coulomb interactions.
We predict a series of electronic spin transitions which arise from the
competition between the many-body gap and magnetic thermal fluctuations.
\end{abstract}
\maketitle

Magnetic doping of semiconductor quantum dots (QDs) provides an interesting
interplay of interaction effects in confined
geometries~\cite{Jacak1998:Book,Reimann2002:RMP,Tarucha1996:PRL,
Mackowski2004:APL,Gould2006:PRL,Leger2006:PRL,Abolfath2006:PRL,Erwin2005:N}
and potential spintronic applications~\cite{Zutic2004:RMP}.
In the bulk-like dilute magnetic semiconductors
the carrier-mediated ferromagnetism can be
photoinduced~\cite{Koshihara1997:PRL,Nunez2004:JMMM} 
and electrically controlled by gate electrodes~\cite{Ohno2000:N},
suggesting possible nonvolatile devices with tunable
optical, electrical, and magnetic properties~\cite{Zutic2004:RMP}.
QDs allow for a versatile
control of the number of carriers, spin, and the effects of quantum
confinement which could lead to improved optical, transport, and magnetic
properties as compared to their bulk
counterparts~\cite{Jacak1998:Book,Holub2004:APL,Erwin2004:NM}. Unlike in the
bulk structures, adding a single carrier in a magnetic QD can have
important ramifications. An extra carrier can both strongly change the
total carrier spin and the temperature of the onset of
magnetization which we show
can be further controlled by modifying the quantum
confinement and the strength of Coulomb interactions.

We study the magnetic ordering of carrier spin and magnetic impurities
in (II,Mn)VI QDs identified as a versatile system to demonstrate
interplay of quantum confinement and magnetism~\cite{Mackowski2004:APL,Gould2006:PRL,
Leger2006:PRL,Hawrylak1991:PRB,Fernandez-Rossier2004:PRL,Govorov2005:PRB,%
Qu2005:PRL}.
Because Mn is isoelectronic with group-II elements it does not change
the number of carriers which in QDs
are controlled by either chemical doping or by external electrostatic
potential applied to the metallic gates. The latter allows confinement of
the carriers in a dot with tunable size and shape~\cite{Reimann2002:RMP}.
By using real space finite-temperature local spin density approximation
(LSDA)~\cite{Dharma-wardana1995:Book}
we study temperature ($T$) evolution of magnetic properties
of QDs over a large parameter space. This approach allow us to consider
QDs with varying number of interacting electrons ($N$) and Mn impurities
($N_m$) which already for small $N$ and $N_m$ becomes computationally
inaccessible to the  exact diagonalization
techniques~\cite{Qu2005:PRL,complexity}.
We extend the previous studies of Coulomb interactions in magnetic QDs
with $N_m=1,2$ at $T=0$~\cite{Qu2005:PRL} and $T>0$ results using
either Thomas-Fermi approximation
or by applying Hund's rule with up 
to 6 carriers~\cite{Govorov2005:PRB}.
We reveal that the interplay of strong Coulomb interactions and quantum
confinement leads to enhanced inhomogeneous magnetization which persist
at higher temperatures than in the non-interacting case
and the bulk structures~\cite{Fernandez-Rossier2004:PRL,Govorov2005:PRB}.
We refer to such a spin-polarized state in QD at zero applied
magnetic field as ``ferromagnetic"
state~\cite{Fernandez-Rossier2004:PRL,Govorov2005:PRB,Qu2005:PRL}.

Here we focus on magnetic QD in zero applied magnetic field
described by the Hamiltonian $H=H_e+H_m+H_{ex}$, with
the electron contribution 
$
H_e = \sum_{i=1}^N 
[-\frac{\hbar^2}{2m^*}\nabla^2_i 
+ U_{QD}({\bf r}_i)]  
+ \frac{\gamma}{\epsilon}\sum_{i\neq j} 
\frac{e^2}{|{\bf r}_i - {\bf r}_j|},
$
where $\hbar$ is the Planck constant, $m^*$ is the electron effective mass, 
and $U_{QD}({\bf r})$ is the confining potential of a three-dimensional QD.
The last term in the equation 
is the repulsive electron-electron (e-e)
Coulomb interaction screened by the dielectric constant $\epsilon$,
$-e$ is electron charge, and $\gamma$ accounts for reduction of Coulomb
strength due to screening effects of the gate 
electrodes~\cite{Bruce2000:PRB}.
The Mn Hamiltonian is
$H_m=\sum_{I,I'}J^{AF}_{I,I'} \vec{M}_I \cdot \vec{M}_{I'}$, where 
$J^{AF}$ is the direct Mn-Mn antiferromagnetic coupling.
The $z$-component of 
$\vec{M}_I$ of impurity spin satisfies $M_z=-M, -M+1, \dots, M$, where we
choose $\hat{z}$ as the quantization axis and $M=5/2$ for Mn.
The electron-Mn (e-Mn) exchange Hamiltonian is
$
H_{ex}=- J_{sd}\sum_{i,I}\vec{s}_i\cdot\vec{M}_I  
\delta({\bf r}_i - {\bf R}_I),
$
where $J_{sd}$ is the exchange coupling between electron spin 
$\vec{s}_i$, at ${\bf r}_i \equiv (\vec{\rho}_i,z_i)$,
and impurity spin $\vec{M}_I$, at ${\bf R}_I$.
An effective mean field Hamiltonian describing electrons can be obtained 
by replacing the Mn spins, that are randomly distributed, with a classical
continuous field
$
H^{\rm eff}_e = H_e - \sum_i J_{sd} n_m \frac{\sigma_i}{2} 
\langle M_z({\bf r}_i)\rangle,
$
where $n_m$ is the averaged density of Mn, and $\sigma=\pm 1$ for 
spin up ($\uparrow$), and down ($\downarrow$).
The effective magnetic field seen by the electrons is the 
mean field induced by Mn.
Assuming that impurities are in equilibrium with thermal bath it follows
$\langle M_z({\bf r}_i)\rangle = M B_M(M b({\bf r}_i)/k_BT)$ where
$B_M(x)$ is the Brillouin function~\cite{Furdyna1988:JAP}, 
$k_B$ is the Boltzmann constant. 
Here $b({\bf r}_i)=- Z_{Mn} J^{AF} \langle M_z({\bf r}_i)\rangle
+ J_{sd} [n_\uparrow({\bf r}_i) - n_\downarrow({\bf r}_i)]/2$
is the effective field seen by the Mn~\cite{DasSarma2003:PRB}. 
The first term in $b({\bf r}_i)$ describes the mean field
of the direct Mn-Mn antiferromagnetic coupling~\cite{Fernandez-Rossier2004:PRL}.
$Z_{Mn}$ is the averaged Mn coordination number, and 
$n_\sigma({\bf r}_i)$ is spin-resolved electron density.
We decompose the planar and perpendicular components of the
confining potential of a single QD, and fit it to
a realistic QD potential~\cite{Kyriakidis2002:PRB}.
The resulting potential, $U_{QD}$, is a sum of a two-dimensional (2D) 
Gaussian 
$V_{QD} = V_0 \exp(-\rho^2/\Delta^2)$ and one-dimensional
parabolic potential $V^z_{QD} = m^*\Omega^2 z^2 / 2$,
where $\vec{\rho}\equiv (x,y)$.
For $V_{QD}$ we find that Gaussian potential is
more realistic than usually studied parabolic potential.
Here $V_0$ and $\Omega$ are the planar depth of the QD minimum, and 
the characteristic subband energy associated with the perpendicular 
confinement.
In typical disk-shaped QDs, and
low density of electrons, only the first subband is filled.
After expanding the QD wave functions in terms of its 
planar $\psi_{i\sigma}(\vec{\rho})$ and subband wave function $\xi(z)$,
we project $H^{\rm eff}_e$  
into a two-dimensional Hamiltonian
by integrating out $\xi(z)$.
In LSDA the two-body Coulomb interaction can be written as 
sum of Hartree potential $V_H$ and 
spin dependent exchange-correlation potential $V^\sigma_{XC}$.
We use Vosko-Wilk-Nusair exchange-correlation functional
\cite{Dharma-wardana1995:Book}, and express the Kohn-Sham Hamiltonian as 
\begin{eqnarray}
H_{KS} = \frac{-\hbar^2}{2m^*} \nabla_\rho^2
+ V_{QD} + \gamma V_{H} + \gamma V^\sigma_{XC}
- \frac{\sigma}{2} h_{sd}(\vec{\rho}),
\label{Heff}
\end{eqnarray}
where 
\begin{eqnarray}
h_{sd}(\vec{\rho}) =  J_{em} \int dz |\xi(z)|^2 
B_M\left( \frac{M b(\vec{\rho}, z)}{k_BT}\right),
\label{hsd}
\end{eqnarray}
and $J_{em} = J_{sd} n_m M$ is the e-Mn exchange coupling. 
The Kohn-Sham eigenvectors and eigenvalues of Eq.~(\ref{Heff}), 
$\psi_{n\sigma}(\vec{\rho})$, and $\epsilon_{n\sigma}$
are calculated numerically.

\begin{figure}
\begin{center}\vspace{1cm}
\includegraphics[width=0.9\linewidth]{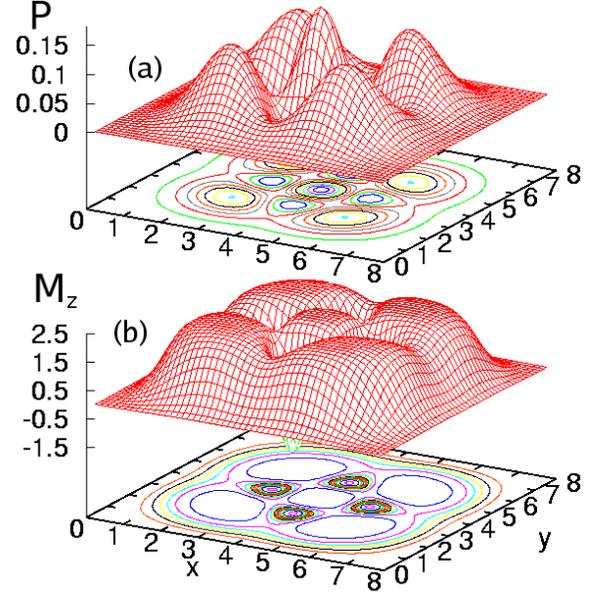}
\caption{
The spatial density profile of electron spin density (a) and 
Mn-magnetization (b) for QD with $N=8$ at $T=1K$.
Coordinates $(x,y)$ are expressed in effective Bohr radius.
}
\label{m8e}
\end{center}
\end{figure}

We illustrate our findings and the iterative solutions of Eq.~(\ref{Heff})
for (Cd,Mn)Te QD.
The material parameters are $J_{sd}=0.015$ eV nm$^3$, $m^*=0.106$, 
$\epsilon=10.6$~\cite{Qu2005:PRL}, and we choose $Z_{Mn} J^{AF} = 0.02$ meV. 
The planar $(x,y)$, and perpendicular ($z$), dimensions of the QD are 
taken as 42 nm and 1 nm with 
$n_m=0$, $0.025$, $0.1$ nm$^{-3}$.
In the central region of QD of area $4a_B^{*2}$, where $a^*_B=5.29$ nm
is the effective Bohr radius in CdTe,
$n_m=0.1$ nm$^{-3}$ corresponds to $\approx 10$ Mn atoms.
For a planar confinement, $V_{QD}$, we consider a Gaussian potential with 
$V_0=-128$ meV, and $\Delta=38.4$ meV, corresponding to 
$\omega_0=27$ meV.
Here $\omega_0$ is calculated by
expanding $V_{QD}$ in the vicinity of the minimum which yields
$V_{QD}=V_0 + m^* \omega_0^2 \rho^2/2 + \dots$, with the strength 
$\omega_0=\sqrt{2|V_0|/m^*} /\Delta$.

In QDs electron density is inhomogeneous, implying that both the
electron spin density, $n_\uparrow(\vec{\rho}) - n_\downarrow(\vec{\rho})$,
and Mn-magnetization density 
$\langle M_z(\vec{\rho}) \rangle \equiv M h_{sd}(\vec{\rho})/J_{em}$ 
are inhomogeneous.
For $N=8$ and $n_m=0.1$ nm$^{-3}$, 
we show the self-consistent spin density in Fig.~\ref{m8e}(a),
and  Mn-magnetization density in Fig.~\ref{m8e}(b). 
Outside the QD, $n_\sigma(\vec{\rho})$ decays exponentially, and 
an effective field $b(\vec{\rho})$ seen by Mn becomes negligible.
This is consistent with vanishing $\langle M_z(\vec{\rho}) \rangle$ at the
QD boundary [Fig.~\ref{m8e}(b)].

\begin{figure}
\begin{center}\vspace{1cm}
\includegraphics[width=0.9\linewidth]{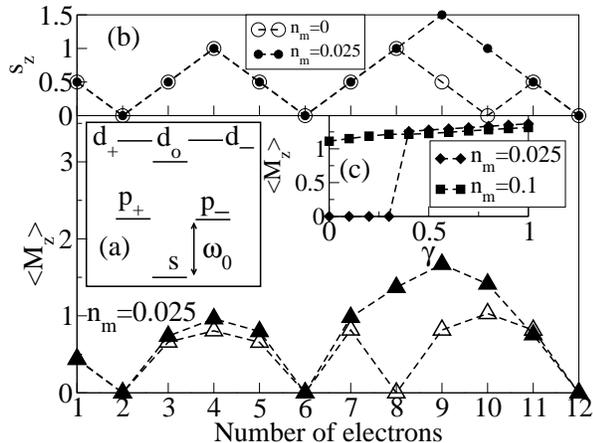}
\caption{
The averaged magnetization per unit area
$\langle M_z\rangle$ as a function of number of electrons
$N$ at $T=1K$ and Mn-density $n_m=0.025$ nm$^{-3}$ for
non-interacting ($\gamma=0$, empty triangles) and 
interacting ($\gamma=1$, filled triangles) electrons.
The ground state of the QD switches between ferromagnetic
and antiferromagnetic states as function of $N$.
Coulomb interaction changes the state of
$N=8$ from antiferromagnetic to ferromagnetic state.
(a) The schematic single particle levels of a 2D 
Gaussian confining potential. 
(b) The $z$-component of the total spin of electrons, 
$s_z$, as a function of $N$ for $\gamma=1$.
(c) Antiferromagnetic-ferromagnetic transitions 
for $N=8$ as function of  $n_m$ and $\gamma$.
}
\label{m_Mn1}
\end{center}
\end{figure}

We next turn to spatially-averaged quantities such as 
Mn-magnetization
per unit area $A$, $\langle M_z \rangle = \frac{1}{A} 
\int d^2\rho \langle M_z(\vec{\rho}) \rangle$, 
electron (spin) 
polarization $P=(N_\uparrow - N_\downarrow)/N$, and
the $z$-component of the total spin of electrons,
$s_z=(N_\uparrow - N_\downarrow)/2$.
In Fig.~\ref{m_Mn1} we show $\langle M_z \rangle$
as a function of $N$ for $n_m=0.025$ nm$^{-3}$, and both 
non-interacting ($\gamma=0$), and 
interacting ($\gamma=1$) electrons. 
The magnetic behavior of QD can be described based on the 
interplay of the many-body spectrum (determined by the
shell structure of the confining potential, and e-e Coulomb
interaction) and the strength of e-Mn exchange coupling 
$J_{em}$. 
In the following we summarize the spin structure of the QD
in the absence and presence of $J_{em}$ with $\gamma=0,1$.

i) $J_{em}=0$: The shell structure of the
2D Gaussian potential is shown in Fig.~\ref{m_Mn1}(a).
The energy gap between $s$-, $p$-, and $d$-orbitals
is characterized by $\omega_0$.
In contrast to 2D parabolic potential~\cite{Jacak1998:Book,Reimann2002:RMP},
$d$-shell levels
are not completely degenerate, and therefore we 
focus on $N>6$ states.
Degenerate levels $d_+$ and $d_-$ are separated by an energy gap
(1.5 meV) from $d_0$-level, 
where $\pm,0$ refer to angular momentum $l_z = \pm 1,0$.
However, e-e interaction 
changes the structure of $d$-shell
as it overturns the ordering of the $d$-orbitals, e.g., 
the Kohn-Sham energies of pair of degenerate $d_+$, and $d_-$ are 
below $d_0$ (with energy gap $\approx$ 1 meV). 
Because of $d$-shell overturning, 
$N=10,12$ form closed shells with $s_z=0$, and $N=7,9,11$ form
open shells with $s_z=1/2$.
The $N=8$ corresponds to half-filled shell with $s_z=1$, and
electron polarization, $P=2/8$.
The evolution of $s_z$ as a function of $N$
for $\gamma=1$ is shown in Fig.~\ref{m_Mn1}(b) (empty circles).

ii) $J_{em} \neq 0$: 
$\gamma=0$ and increasing e-Mn coupling to $J_{em}= 3.75$ meV,
leads to transitions $P=0/8\rightarrow 2/8$, and
$P=1/9\rightarrow 3/9$ for $N=8$, and $9$,
whereas $N=10$ shows $P=2/10$.
For $\gamma=1$, the dependence of $P$ on $J_{em}$
is negligible in $s$-, and $p$- shells.
In contrast, in $d$-shell, we find transitions  
$P=1/9 \rightarrow 3/9$, and $P=0/10 \rightarrow 2/10$
for $N=9$ and $N=10$ at low T.
However, we find no change in $P$ for $N=7,8,11$, and 12.
Figure~\ref{m_Mn1}(b) (filled circles) show $s_z$ as a function of $N$
for $n_m=0.025$ nm$^{-3}$ ($J_{em}=0.94$ meV).
Increasing the density of Mn to $n_m=0.1$ nm$^{-3}$ ($J_{em}=3.75$ meV)
does not change $s_z$.

In Fig.~\ref{m_Mn1}, we observe that $\langle M_z \rangle = 0$ in 
closed shells ($N=2,6,12$) for both $\gamma=0$ and $\gamma=1$, 
because of well separated 
$s$-, $p$-, and $d$- orbitals due to large
$\omega_0 (\approx 30 J_{em})$.
Comparing $\langle M_z \rangle$ between $\gamma=0$
and $\gamma=1$ one can observe that
the e-e interaction stabilizes the ferromagnetic state 
due to the spin Hund's rule.
This condition is easily satisfied for open shells where the maximum 
electron polarization is obtained in half-filled shell with $N=4$.
The $N=8$  state (recall Fig.~\ref{m8e}) is more interesting. 
At $\gamma=0$ electrons fill single particle levels 
following the Pauli exclusion principle.
Even with $J_{em} = 0.94$ meV (smaller than single particle e-h excitation
gap), $N=8$ forms closed shell and $P=\langle M_z \rangle=0$.
In the case of $\gamma=1$, and because of $d$-shell overturning,
polarized electrons in $d_+$ and $d_-$ give $P=2/8$, and 
finite $\langle M_z \rangle$.
We also see that the maximum $\langle M_z \rangle$ 
occurs at $N=9$ because $J_{em} = 0.94$ meV induces 
three polarized electrons in $d$-levels.
Figure \ref{m_Mn1}(c) reveals the dependence of magnetic transitions
on $\gamma$, and $n_m$.
With increasing $Z_{Mn} J^{AF}$, the transition to ferromagnetic state
occurs at larger $\gamma$.
Our findings clearly demonstrate that the magnetism induced by
strong Coulomb interaction 
can be controlled by the electric gates
or by changing the semiconductor host
(and thus changing $\epsilon$)
without changing the number of carriers confined in QD.
We also suggest that because of the sensitivity of 
the $\langle M_z \rangle$ to the electronic 
spin transitions, the former can be used to infer the spin 
of electrons, and could be potentially applied to manipulation of spin 
qubits in semiconductor nanostructures~\cite{Zutic2004:RMP}.

\begin{figure}
\begin{center}\vspace{1cm}
\includegraphics[width=0.9\linewidth]{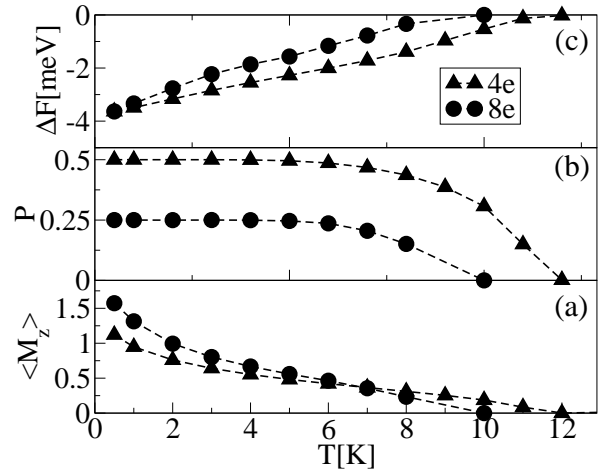}\vspace{1cm}
\caption{
Temperature evolution of Mn-magnetization per unit area
$\langle M_z \rangle$ (a), the electron polarization $P$ (b), and
the free energy difference $\Delta F$ 
between ferromagnetic and antiferromagnetic QD (c).
At low T, $N=4,8$ form half-filled shells with $P=2/4,2/8$.
$T=T^*$, characterizes vanishing of 
$\langle M_z \rangle$, $P$, and $\Delta F$.
}
\label{Mag_Pol_F}
\end{center}
\end{figure}

We next examine the temperature dependence of magnetism in QDs.
In Fig.~\ref{Mag_Pol_F}, we show $\langle M_z\rangle$ (a), $P$ (b), and
the free energy difference $\Delta F$ between ferromagnetic
and antiferromagnetic states (c) for $N=4$ and $N=8$.
The suppression of $\langle M_z \rangle$, shown 
in Fig.~\ref{Mag_Pol_F}(a) is accompanied by a series of 
spin transitions in electronic states and suppression of $P$.
At low T, the spin triplet 
is realized as the ground state of the $N=4$, and $N=8$ open 
$p$-, and $d$-shells (due to Hund's rule).
We define a characteristic temperature, $T^*$, 
at which $\langle M_z \rangle=P=\Delta F=0$.

\begin{figure}
\begin{center}\vspace{1cm}
\includegraphics[width=0.9\linewidth]{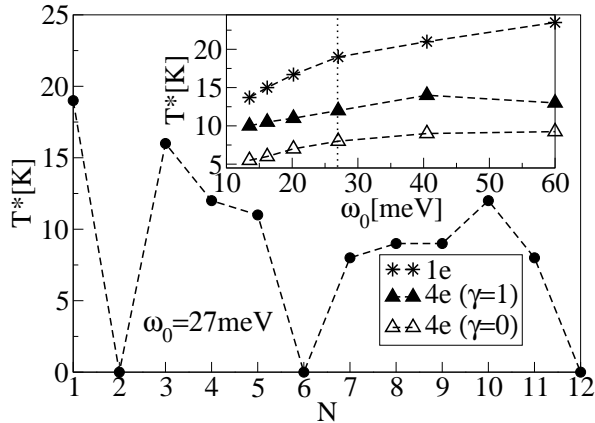}\vspace{1cm}
\caption{
$T^*$ as a function of $N$ for interacting electrons $\gamma=1$,  
$\omega_0=27$ meV, and $V_0=-125$ meV.
Inset: The dependence of $T^*$ on $\omega_0$
for $N=1$, $N=4$ with $\gamma=0$, and $\gamma=1$.
$\omega_0=27$ meV is marked as a dotted line.
There is an optimal confining potential which maximizes $T^*$.
}
\label{Tc}
\end{center}
\end{figure}

In Fig.~\ref{Tc} we plot $T^*(N)$ for $\omega_0=27$ meV and
$\gamma=1$ which decreases non-monotonically with $N$. 
The inset shows $T^*(\omega_0)$ for $N=1$, and $N=4$ ($\gamma=0,1$).
At low $\omega_0$, the e-e interaction strongly enhances $T^*$,
while at large $\omega_0$ the effect of confinement potential 
is dominant. 
Thus we find $T^*(\gamma=1) \rightarrow T^*(\gamma=0)$ 
with increasing $\omega_0$,
which in turn gives rise to a peak in $T^*$.
Several trends in calculated $T^*(N,\omega_0)$ can be obtained
from a perturbative approach by approximating 2D Gaussian 
with 2D parabolic potential.
Near $\langle M_z \rangle =P=0$ for QD with one valence electron
in $s$-, $p$-, or $d$-shells, we find 
$T^* = J_{em} \sqrt{\frac{M+1}{3 n_m M}} \left[
\int d^3 {\bf r} |\psi_f({\bf r})|^4\right]^{1/2}$, where
$\psi_f$ is the wave-function of the
highest occupied orbital, and $J^{AF} = 0$.
For a given $\omega_0$,  
$T^*$ decreases with $N$, e.g., 
$T^*_{N=3} = 0.7 T^*_{N=1}$ and  $T^*_{N=7} = 0.6 T^*_{N=1}$.
One can also show that $T^* \propto \sqrt{\omega_0}$,
consistent with bound magnetic polarons~\cite{Schmidt1999:PRL}.

In conclusion we have investigated the existence of magnetism in
magnetically doped
QDs, as function of particle numbers, confining potential,
temperature and strength of Coulomb interactions, using
finite-temperature LSDA. 
Our results show that 
QDs embedded in magnetic semiconductor host
can be considered as ferromagnetic centers which exhibit spatial
ordering in spin density and magnetization,
even at elevated temperatures where no such ordering
exist in the host material~\cite{Dietl1997:PRB}.
In the limit of small $\omega_0$,
where the Coulomb interaction among particles
is the largest characteristic parameter of the QDs, we
find magnetism substantially stronger than predicted 
from the non-interacting picture.
In contrast to the carrier-controlled ferromagnetism in the bulk-like
structures~\cite{Koshihara1997:PRL,Nunez2004:JMMM,Ohno2000:N}, we reveal 
that magnetism in QDs, can be tuned
even at the fixed number of carriers by gate voltage which
controls the inter-particle Coulomb interaction screening.
For potential spintronic applications based on II-VI magnetic QDs,
we anticipate that it is possible to further increase the
magnetization and the temperature at which it vanishes. In addition
to exploring  a larger hole-Mn exchange coupling~\cite{Furdyna1988:JAP} 
in (II,Mn)VI QDs, it would also be advantageous to consider (II,Cr)VI
QDs as there is 
a support for the room-temperature ferromagnetism in 
their bulk counterparts~\cite{Saito2003:PRL}.

This work is supported by the US ONR, NSF-ECCS Career, the NRC HPC project,
CIAR, and the CCR at SUNY Buffalo.
We thank C. Dharma-wardana for stimulating discussions.

       



\begin{thebibliography}{99}

\bibitem{Jacak1998:Book}
L. Jacak, P. Hawrylak, and A. Wojs, {\em Quantum Dots} (Springer, Berlin, 1998);
D. Bimberg, M. Grundmann, N.N. Ledentsov, {\em Quantum Dot Heterostructures},
John Wiley \& Sons (Chichester 1999).

\bibitem{Reimann2002:RMP}
S. M. Reimann and M. Manninen, Rev. Mod. Phys. {\bf 74}, 1283 (2002),
and the references therein.

\bibitem{Tarucha1996:PRL}
S. Tarucha, D. G. Austing, and T. Honda, R. J. van der Hage and 
L. P. Kouwenhoven, Phys. Rev. Lett. {\bf 77}, 3613 (1996).

\bibitem{Mackowski2004:APL}
S. Mackowski, T. Gurung, T. A. Nguyen, H. E. Jackson, and L. M. Smith,
G. Karczewski and J. Kossut, \apl {\bf 84}, 3337 (2004).

\bibitem{Gould2006:PRL} 
C. Gould, A. Slobodskyy, T. Slobodskyy, P. Grabs, D. Supp, P. Hawrylak, 
F. Qu, G. Schmidt, L.W. Molenkamp, \prl {\bf 97}, 017202 (2006).

\bibitem{Leger2006:PRL} 
Y. L\'eger, L. Besombes, J. Fern\'andez-Rossier, L. Maingault, and H. Mariette,
Phys. Rev. Lett. {\bf 97}, 107401 (2006);
Y. L\'eger, L. Besombes, L. Maingault, D. Ferrand, and H. Mariette,
Phys. Rev. Lett.  {\bf 95}, 047403 (2005).

\bibitem{Abolfath2006:PRL} 
R. M. Abolfath and P. Hawrylak, \prl 97, 186802 (2006).

\bibitem{Erwin2005:N}
S.~C. Erwin , Lijun Zu, Michael I. Haftel, Alexander L. Efros, 
Thomas A. Kennedy, David J. Norris, Nature {\bf 436}, 91 (2005).

\bibitem{Zutic2004:RMP} 
I. \v{Z}uti\'c, J. Fabian, and S. Das Sarma, Rev. Mod. Phys. {\bf 76}, 
323 (2004).
 
\bibitem{Koshihara1997:PRL}
S. Koshihara, A. Oiwa, M. Hirasawa, S. Katsumoto, Y. Iye, C. Urano, 
H. Takagi, and H. Munekata, \prl {\bf 78}, 4617 (1997).

\bibitem{Nunez2004:JMMM}
Alvaro S. Nunez, J. Fernndez-Rossier, M. Abolfath, and A. H. MacDonald,
J. Magn. Magn. Mater. {\bf 272}, 1913 (2004).
 
\bibitem{Ohno2000:N}
H. Ohno, D. Chiba, F. Matsukura, T. Omiya, E. Abe, T. Dietl, 
Y. Ohno, K. Ohtani, Nature (London) {\bf 408}, 944 (2000);
D. Chiba, M. Yamanouchi, F. Matsukura, and H. Ohno,
Science {\bf 301}, 943 (2003).

\bibitem{Holub2004:APL}
M. Holub, S. Chakrabarti, S. Fathpour, and P. Bhattacharya,
Y. Lei, and S. Ghosh,  \apl {\bf 85}, 973 (2004).

\bibitem{Erwin2004:NM}
S.~C. Erwin and I. \v{Z}uti\'c, Nature Mater. {\bf 3}, 410 (2004).

\bibitem{Hawrylak1991:PRB}
P. Hawrylak, M. Grabowski and J.J. Quinn, 
Phys. Rev. B {\bf 44}, 13082 (1991).

\bibitem{Fernandez-Rossier2004:PRL} 
J. Fern\'andez-Rossier and L. Brey, Phys. Rev. Lett. {\bf 93}, 117201 (2004).
J. Fern\'andez-Rossier, Phys. Rev. B {\bf 73}, 045301 (2006).

\bibitem{Govorov2005:PRB} 
A. O. Govorov
Phys. Rev. B {\bf 72}, 075358 (2005);
Phys. Rev. B {\bf 72}, 075359 (2005).

\bibitem{Qu2005:PRL} 
F. Qu and P. Hawrylak,
Phys. Rev. Lett. {\bf 95}, 217206 (2005); 
Phys. Rev. Lett. {\bf 96}, 157201 (2006).

\bibitem{Dharma-wardana1995:Book}
C. Dharma-wardana, and F. Perrot in {\em Density Functional Theory}, 
Edited by E.~K.~U. Gross, and R.M. Dreizler (Plenum Press, New York, 1995).

\bibitem{complexity}
The size of the Hamiltonian matrix to be diagonalized for just
10 Mn (spin-5/2) is $6^{10} \approx 6 \times 10^7$.

\bibitem{Bruce2000:PRB}
N. A. Bruce and P. A. Maksym, Phys. Rev. B {\bf 61}, 4718 (2000).

\bibitem{Furdyna1988:JAP}
J. K. Furdyna, J. Appl. Phys. {\bf 64}, R29 (1988).

\bibitem{DasSarma2003:PRB} 
S. Das Sarma, E. H. Hwang, and A. Kaminski, 
Phys. Rev. B {\bf 67}, 155201 (2003).

\bibitem{Kyriakidis2002:PRB} 
J. Kyriakidis, M. Pioro-Ladriere, M. Ciorga, A. S. Sachrajda, and P. Hawrylak,
Phys. Rev. B {\bf 66}, 035320 (2002).

\bibitem{Schmidt1999:PRL}
D. R. Schmidt, A. G. Petukhov, M. Foygel, J. P. Ibbetson, and S. J. Allen,
Phys. Rev. Lett. {\bf 82}, 823 (1999).

\bibitem{Dietl1997:PRB}
T. Dietl, A. Haury and Y. Merle d'Aubign\'e,  
Phys. Rev. B {\bf 55}, R3347 (1997).

\bibitem{Saito2003:PRL}
H. Saito, V. Zayets, S. Yamagata, and K. Ando, \prl {\bf 90}, 207202 (2003).  

\end{thebibliography}
\end{document}